\documentclass[%
 reprint,
 superscriptaddress,
 amsmath,amssymb,
 aps,
 pra,
]{revtex4-2}

\usepackage{graphicx}% Include figure files
\usepackage{dcolumn}% Align table columns on decimal point
\usepackage{bm}% bold math
\usepackage{amsmath}
\usepackage{braket}
\usepackage{xcolor}
\usepackage{comment}
\usepackage{siunitx}
\begin{document}

\preprint{APS/123-QED}

\title{
Bogoliubov analysis of Higgs mode \\
in trapped Fermi superfluids  with spatial inhomogeneity
}% Force line breaks with \\
%\thanks{A footnote to the article title}

\author{K. Kakimoto}
\affiliation{%
Department of Electronic and Physical Systems,
Waseda University, Tokyo 169-8555, Japan
}%
\author{J. Takahashi}
\affiliation{%
Department of Economics, Asia University,
Tokyo 180-8629, Japan
}%
\author{Y. Yamanaka}
\affiliation{%
Institute of Condensed-Matter Science,
Waseda University, Tokyo 162-0041, Japan
}%

\date{\today}% It is always \today, today,

\begin{abstract}
\begin{comment}
The spontaneous breaking of a continuous symmetry incorporates
Higgs mode as collective amplitude oscillations of order parameters.
\textcolor{red}{
Recently, ultracold atomic gases have attracted considerable attention
as new experimental platforms to explore the properties of
the Higgs mode due to their high controllability
to realize box-shaped trap.
At the same time, they are inherently inhomogeneous
due to the confinement of neutral atoms in trapping potentials.
The Higgs mode has been extensively studied in homogeneous systems,
while its properties in inhomogeneous systems remain elusive,
and a general steady-state framework applicable to
a wide range of trap potentials has yet to be established.}
In this paper,
we investigate the Higgs mode in harmonically trapped Fermi superfluids.
Within the Hatree-Fock approximation, we derive integral equations from
the Bogoliubov-de Gennes equations, which lead to the frequencies of the
collective modes, including the Higgs and Nambu-Goldstone modes.
The results show that the frequency of the Higgs mode equals twice
the absolute value of the order parameter at the center of trap.
This feature is robust against variations in the interaction strength,
trap potential, and temperature.
These results are consistent with previous theoretical and
experimental studies.
\end{comment}
The Higgs mode is a key component in the spontaneous breaking of
a continuous symmetry along with the Nambu-Goldstone mode,
and has been studied extensively for homogeneous systems.
We consider it for inhomogeneous systems,
using the superfluid of harmonically trapped ultracold Fermi atomic gas.
The Fermionic field operators are expanded in a complete set of
wave functions corresponding to inhomogeneous situation.
Within the Hatree-Fock approximation,
we derive integral equations from the Bogoliubov-de Gennes equations,
which lead to the frequencies of the collective modes,
including the Higgs and Nambu-Goldstone modes.
The results show that the frequency of the Higgs mode equals twice
the absolute value of the order parameter at the center of trap.
This feature is robust against variations in the interaction strength,
trap potential, and temperature.
These results are consistent with previous theoretical
and experimental studies.

%\begin{description}
%You may use the \texttt{description} environment to %structure your abstract;
%use the optional argument of the \verb+\item+ command to %give the category of each item. 
%\end{description}
\end{abstract}

\maketitle

%\tableofcontents

\section{\label{sec:level1}INTRODUCTION
%:\protect\\ The line
%break was forced \lowercase{via}
}

The spontaneous breaking of the global $U(1)$ symmetry
gives rise to Higgs and Nambu–Goldstone (NG) modes in
a variety of systems~\cite{Nambu1960, Nambu1961-1, Nambu1961-2, Goldstone1961, Englert1964, Higgs1964, Guralnik1964}.
These two collective modes play a fundamental role in physics over
a wide range of energy scales,
from condensed matter physics~\cite{Pekker2015, Beekman2019, Shimano2020, Tsuji2024}
(low energy)
to elementary particle physics~\cite{Dittmaier2013, Panico2016, Bass2021}
and cosmology \cite{Moss2015, Horn2020, Lebedev2021} (high energy).
This implies that
revealing the properties of the two modes
contributes not only to a specific field
but also to various branches of physics.
In particular, the Higgs mode in table-top systems,
such as ultracold atomic gases, is expected to
offer valuable insights into spontaneous symmetry breaking
and collective excitations.
This is because it is feasible to observe the Higgs mode
using optical techniques under controllable environments
with spontaneous U(1) symmetry breaking
and we can explore the properties of the Higgs mode in detail.

In condensed matter physics including ultracold atomic gas,
the Higgs mode has recently been observed in a broad class of systems,
such as superconductors
\cite{Matsunaga2013, Matsunaga2014, Measson2014, Sherman2015, Kastumi2018},
charge density wave \cite{Cea2014, Wang2022, Wulferding2024},
antiferromagnets \cite{Ruegg2008, Jain2017, Hong2017, Souliou2017},
bosonic superfluids in optical lattices
\cite{Bissbort2011, Endres2012, Leonard2017},
Rydberg atoms~\cite{Liu2024},
and ultracold Fermi gases \cite{Behrle2018, Dyke2024, Kell2024, Breyer2025}.
These experiments were conducted in homogeneous systems
or regions where a local density approximation is valid.
The experimental realization of (quasi-)uniform systems allows us
high-precision comparisons with theories,
because approximate analytical solutions for the uniform systems
are known in most cases.
The key properties of homogeneous Fermi superfluids are well understood,
i.e. the energy gap $\Delta$ suppresses low-energy single-particle
excitations, and the minimum excitation energy is given by $2|\Delta|$,
which equals the frequency of the Higgs mode
\cite{Anderson1958, Littlewood1981, Littlewood1982, Kulik1981, Tsuji2015, Volkov1974}.

In this paper, we study the Higgs mode in ultracold atomic systems,
focusing on the inhomogeneity.
Since these systems are confined by external traps,
the resulting spatially varying density may significantly affect
the properties of the Higgs mode.
While our study of the Higgs mode in inhomogeneous systems have primarily
focused on harmonically trapped gases, the theoretical frameworks
are applicable to a broad class of trapping potentials
\cite{Scott2012, Tokimoto2017, Tokimoto2019, Hannibal2018, Bruun2014, Dutta2017, Bjerlin2016}.
Because many physical systems in nature are inherently inhomogeneous,
theoretical frameworks for such systems contribute
to a detailed understanding of the Higgs mode
not only in condensed matter systems but also in cosmology,
elementary particle and nuclear physics~\cite{Takahashi2023}.

The studies of the Higgs mode in inhomogeneous systems
can be categorized into two approaches.
One is a dynamical approach
that involves solving time evolution equations
to directly induce oscillations of the order parameter
\cite{Scott2012, Tokimoto2017, Tokimoto2019, Hannibal2018}.
The other is a static approach that determines
the collective modes by solving a matrix equation
\cite{Bruun2014, Dutta2017}.
In the latter approach, amplitude and phase oscillations
correspond directly to the Higgs and NG modes, respectively. Moreover,
additional collective modes can be obtained through this approach
\cite{Dutta2017}.
This implies that static approaches have the potential to unveil
microscopic structures of the collective modes.
Our study in this paper belongs to the static approach.

\begin{comment}
Although these theoretical studies focus on specific systems,
such as harmonic trap systems,
these theoretical frameworks can be applied to a wide range of systems
featuring spatial inhomogeneity,
including the boundary conditions of box traps \cite{Liebster2025},
finite-size effects in small systems \cite{Reible2023}, and
background structures~\cite{Soto-Garrido2017}.
In particular,
the Fulde–Ferrell–Larkin–Ovchinnikov (FFLO) state naturally
involves the spatial dependence of the order parameter
due to Cooper pairs with finite center-of-mass momentum
~\cite{Dutta2017, Casalbuoni2004, Matsuda2007, Shimahara2007, Flude1964, Larkin1964, Zhao2022, Zou2024, Miwa2025}.

In this paper,
\end{comment}

Specifically, we present the formulation for a trapped
superfluid Fermi gas. The field operators are expanded 
in an appropriate complete set of wave functions
corresponding to inhomogeneous situation. It is stressed that
our approach is based on quantum field theory.
Aiming for investigating the Higgs mode in the inhomogeneous superfluid
system, we derive integral equations,
which provide the frequencies of collective
excitations, including the Higgs and NG modes. The integral
equations reproduce, in the homogeneous limit, 
the frequencies of the two modes,
obtained analytically in homogeneous systems.
In addition, we perform numerical calculations to determine
the frequency of the Higgs mode in inhomogeneous
systems confined to harmonic traps.

The paper is organized as follows.
Section \ref{formulation} derives integral equations
describing the Higgs and NG modes in trapped Fermi
superfluids within the Hartree-Fock approximation.
In Section \ref{solve_equation},
we solve the integral equation
in homogeneous and inhomogeneous systems.
Section \ref{conclusions} summarizes the conclusions.

\section{System and formulation}
\label{formulation}
We investigate a trapped three-dimensional superfluid
Fermi gas with equal populations of
the two spin components, described by the Hamiltonian
\begin{equation}
    \hat{H}
    =   \int d^3\mathbf{r}
        \sum_{\sigma=\uparrow, \downarrow}
        \hat{\psi}_{\sigma}^{\dagger}
        H_0 \hat{\psi}_{\sigma}
        -
        g \int d^3\mathbf{r} \,
        \hat{\psi}_{\uparrow}^{\dagger}
        \hat{\psi}_{\downarrow}^{\dagger}
        \hat{\psi}_{\downarrow} \hat{\psi}_{\uparrow},
\end{equation}
where $\uparrow(\downarrow), \mathbf{r}$, $t$ and $g$ are
the up (down) spin, position in space, time
and coupling constant, respectively.
The Fermi field operators
$\hat{\psi}_{\sigma} \equiv \hat{\psi}_{\sigma}(\mathbf{r}, t)$
satisfy the following anticommutation relations
\begin{align}
    \left\{
        \hat{\psi}_{\sigma}(\mathbf{r}, t),
        \hat{\psi}_{\sigma'}^{\dagger}(\mathbf{r}', t)
    \right\}
    &=  \delta_{\sigma\sigma'}
        \delta(\mathbf{r}-\mathbf{r}'), \\
    \left\{
        \hat{\psi}_{\sigma}(\mathbf{r}, t),
        \hat{\psi}_{\sigma'}(\mathbf{r}', t)
    \right\}
    &=
    \left\{
        \hat{\psi}_{\sigma}^{\dagger}(\mathbf{r}, t),
        \hat{\psi}_{\sigma'}^{\dagger}(\mathbf{r}', t)
    \right\}
    =   0.
\end{align}
The single-particle Hamiltonian is
$H_0=-\hbar^2\nabla^2/2m+V_{\text{trap}}(\mathbf{r})-\mu$,
where $m, V_{\text{trap}}(\mathbf{r})$ and $\mu$ are
the atomic mass, trapping potential and
chemical potential, respectively.

We perform the Hartree-Fock approximation with
the superfluid order parameter, and decompose ${\hat H}$
into the unperturbed and perturbed Hamiltonians,
denoted by ${\hat H}_{\rm u}$ and ${\hat H}_{\rm p}$, respectively,
\begin{equation}
    \hat{H} = \hat{H}_u + \hat{H}_p,
\end{equation}
\begin{equation}
    \hat{H}_u
    =   \int d^3\mathbf{r}
        \left[ \begin{matrix}
            \hat{\psi}_{\uparrow}^{\dagger} &
            \hat{\psi}_{\downarrow}
        \end{matrix} \right]
        \left[ \begin{matrix}
            \mathcal{L}(x) & \Delta(x) \\
            \Delta^*(x) & -\mathcal{L}(x)
        \end{matrix} \right]
        \left[ \begin{matrix}
            \hat{\psi}_{\uparrow} \\
            \hat{\psi}_{\downarrow}^{\dagger}
        \end{matrix} \right],
\end{equation}
\begin{equation}
\begin{split}
    \hat{H}_p
    &=  -g \int d^3\mathbf{r} \,
        \hat{\psi}_{\uparrow}^{\dagger}
        \hat{\psi}_{\downarrow}^{\dagger}
        \hat{\psi}_{\downarrow} \hat{\psi}_{\uparrow} \\
        &\quad+
        \int d^3\mathbf{r}
        \left[ \begin{matrix}
            \hat{\psi}_{\uparrow}^{\dagger} &
            \hat{\psi}_{\downarrow}
        \end{matrix} \right]
        \left[ \begin{matrix}
            gn(x) & -\Delta(x) \\
            -\Delta^*(x) & -gn(x)
        \end{matrix} \right]
        \left[ \begin{matrix}
            \hat{\psi}_{\uparrow} \\
            \hat{\psi}_{\downarrow}^{\dagger}
        \end{matrix} \right].
\end{split}
\end{equation}
Here $x$ stands for $({\bm x},t)$,
the order parameter $\Delta(x)$ and the density function $n(x)$
are defined as
\begin{align}
    \Delta(x)
    &=  -g\Braket{
            \hat{\psi}_{\downarrow}(x)
            \hat{\psi}_{\uparrow}(x)
        }, \\
    n(x)
    &=  \Braket{
            \hat{\psi}_{\uparrow}^{\dagger}(x)
            \hat{\psi}_{\uparrow}(x)
        }
    =   \Braket{
            \hat{\psi}_{\downarrow}^{\dagger}(x)
            \hat{\psi}_{\downarrow}(x)
        },
\end{align}
and the symbol ${\mathcal L}(x)$ is given by
\begin{align}
{\mathcal L}(x)=H_0-gn(x)\,.
\end{align}
In the interaction picture, the unperturbed
Hamiltonian leads to the Heisenberg equation
\begin{equation}
    i\hbar\frac{\partial}{\partial t}
    \left[ \begin{matrix}
        \hat{\psi}_{\uparrow} \\
        \hat{\psi}_{\downarrow}^{\dagger}
    \end{matrix} \right]
    =
    \left[ \begin{matrix}
            \mathcal{L}(x) & \Delta(x) \\
            \Delta^*(x) & -\mathcal{L}(x)
        \end{matrix} \right]
    \left[ \begin{matrix}
        \hat{\psi}_{\uparrow} \\
        \hat{\psi}_{\downarrow}^{\dagger}
    \end{matrix} \right].
\label{Heisenberg}
\end{equation}

In a neutral Fermi gas, the normal-superfluid phase transition is
associated with the spontaneous breaking of the global $U(1)$ symmetry.
\begin{comment}
%To visually explain this phenomenon
%in an easily understandable way,
%the Mexican hat potential is commonly used.
%At high temperatures, a free energy surface
%in the plane of a complex order parameter $\Delta$
%becomes a monotonically increasing function of the
%amplitude $|\Delta|$. The free energy attains
%the minimum value at $\Delta=0$ and the system
%takes the normal fluid state.
%At low temperatures, the free energy surface forms
%a Mexican hat (See Fig. \ref{Mexican_Hat}).
%The states with the lowest free energy,
%which have different phases $\theta$ of
%the order parameter $\Delta=|\Delta|e^{i\theta}$, 
%exist at finite amplitude $|\Delta|$.
%When the system becomes the superfluid state,
%one of the states with a specific phase $\theta$,
%belonging to the minimum free energy,
%is spontaneously selected.
%Consequently, the global $U(1)$ symmetry is broken.
%\begin{figure}
%    \includegraphics[width=\linewidth]
%   {MaxicanHat.png}% Here is how to import EPS art
%   \caption{
%        Conceptual diagram  of a Mexican hat potential.
%   }
%\label{Mexican_Hat}
%\end{figure}
In a Fermi superfluid system,
the fluctuations around the steady state
lead to the two bosonic collective modes.
The amplitude and phase oscillations cause
the Higgs and NG modes, respectively~\cite{Pekker2015}.
We consider a situation where
a Fermi superfluid is slightly
deviated from the steady state.
\end{comment}
The fluctuations around the steady state of Fermi superfluid
give rise to the two bosonic collective modes.
They are the Higgs and NG modes, emerging as the amplitude and phase
oscillations, respectively \cite{Pekker2015, Beekman2019, Shimano2020, Tsuji2024}.
We therefore consider a situation where the steady state of Fermi superfluid is perturbed.

\subsection{Steady state}

Assuming that the system is in a steady state and
the physical quantities are time-independent,
we obtain the eigenvalue problem
\begin{equation}
    \left[ \begin{matrix}
            \mathcal{L}^{(0)}(\mathbf{r}) &
            \Delta^{(0)}(\mathbf{r}) \\
            \Delta^{(0)*}(\mathbf{r}) &
            -\mathcal{L}^{(0)}(\mathbf{r})
        \end{matrix} \right]
    \left[ \begin{matrix}
        u_{\nu}^{(0)}(\mathbf{r}) \\
        v_{\nu}^{(0)}(\mathbf{r})
    \end{matrix} \right]
    =
    E_{\nu}
    \left[ \begin{matrix}
        u_{\nu}^{(0)}(\mathbf{r}) \\
        v_{\nu}^{(0)}(\mathbf{r})
    \end{matrix} \right],
\label{stedy_BdG}
\end{equation}
where
\begin{align}
    \mathcal{L}^{(0)}(\mathbf{r})
    &=  H_0 - gn^{(0)}(\mathbf{r}).
\end{align}
Due to the Hermitian property,
the eigenvalue $E_{\nu}$ is real.
Here, the superscript $(0)$
represents the stationary solutions.
Equation (\ref{stedy_BdG}),
which is known as
the time-independent Bogoliubov-de Gennes (BdG) equations,
has a conjugate solution for each eigenvalue $E_{\nu}$,
which stems from the particle-hole symmetry,
\begin{equation}
    \left[ \begin{matrix}
            \mathcal{L}^{(0)}(\mathbf{r}) &
            \Delta^{(0)}(\mathbf{r}) \\
            \Delta^{(0)*}(\mathbf{r}) &
            -\mathcal{L}^{(0)}(\mathbf{r})
        \end{matrix} \right]
    \left[ \begin{matrix}
        -v_{\nu}^{(0)*}(\mathbf{r}) \\
        u_{\nu}^{(0)*}(\mathbf{r})
    \end{matrix} \right]
    =
    -E_{\nu}
    \left[ \begin{matrix}
        -v_{\nu}^{(0)*}(\mathbf{r}) \\
        u_{\nu}^{(0)*}(\mathbf{r})
    \end{matrix} \right].
\end{equation}
In what follows, we restrict our analysis to
the positive eigenvalues $E_{\nu}$,
as the negative eigenvalues correspond to
conjugate solutions.

We expand the Fermi field operators in terms of
the eigenfunctions
\begin{equation}
    \left[ \begin{matrix}
        \hat{\psi}_{\uparrow}(x) \\
        \hat{\psi}_{\downarrow}^{\dagger}(x)
    \end{matrix} \right]
    =
    \sum_{\nu>0}
    \left(
    \left[ \begin{matrix}
        u_{\nu}^{(0)}(\mathbf{r}) \\
        v_{\nu}^{(0)}(\mathbf{r})
    \end{matrix} \right]
    \hat{A}_{\nu}(t)
    +
    \left[ \begin{matrix}
        -v_{\nu}^{(0)*}(\mathbf{r}) \\
        u_{\nu}^{(0)*}(\mathbf{r})
    \end{matrix} \right]
    \hat{B}_{\nu}^{\dagger}(t)
    \right),
\end{equation}
where $\sum_{\nu>0}$ denots the sum over all
positive integers.
The annihilation operators
$\hat{A}_{\nu}(t)$ and $\hat{B}_{\nu}(t)$ represent
the quasiparticles and quasiholes, respectively.
For simplicity,
we associate the negative eigenvalues,
their corresponding eigenvectors, and
the quasiholes with negative integers
\begin{equation}
    \left[ \begin{matrix}
        u_{-\nu}^{(0)} \\
        v_{-\nu}^{(0)}
    \end{matrix} \right]
    =
    \left[ \begin{matrix}
        -v_{\nu}^{(0)*} \\
        u_{\nu}^{(0)*}
    \end{matrix} \right]
    , \quad
    E_{-\nu} = -E_{\nu}
    , \quad
    \hat{A}_{-\nu}(t) = \hat{B}_{\nu}^{\dagger}(t).
\end{equation}
The Fermi operator $\hat{A}_{\nu}(t)$ satisfies the
anticommutation relations
\begin{align}
    \left\{
        \hat{A}_{\mu}(t), \hat{A}_{\nu}^{\dagger}(t)
    \right\}
    &=
    \delta_{\mu\nu}, \\
    \left\{
        \hat{A}_{\mu}(t), \hat{A}_{\nu}(t)
    \right\}
    &=
    \left\{
        \hat{A}_{\mu}^{\dagger}(t), \hat{A}_{\nu}^{\dagger}(t)
    \right\}
    =   0.
\end{align}
where $\mu$ and $\nu$ represent certain nonzero
integers.
Using the orthonormality of the eigenfunctions
\begin{equation}
    \int d^3\mathbf{r}
    \left(
        u_{\mu}^{(0)*} u_{\nu}^{(0)}
        +
        v_{\mu}^{(0)*} v_{\nu}^{(0)}
    \right)
    =   \delta_{\mu\nu},
\label{orthonormality}
\end{equation}
we obtain
\begin{equation}
    \hat{H}_u
    =   \sum_{\nu}
        E_{\nu}
        \hat{A}_{\nu}^{\dagger}(t)
        \hat{A}_{\nu}(t),
\end{equation}
where $\sum_{\nu}$ denotes the sum over all
nonzero integers.
From the Heisenberg equation
\begin{align}
    i\hbar\frac{\partial}{\partial t}
    \hat{A}_{\nu}(t)
    &=
    \left[
        \hat{A}_{\nu}(t), \hat{H}_u
    \right]
    =   E_{\nu} \hat{A}_{\nu}(t),
\end{align}
we obtain
$\hat{A}_{\nu}(t)
=e^{-iE_{\nu}t/\hbar}\hat{A}_{\nu}(0)
$.
The Fermi field operators are expanded as
\begin{equation}
    \left[ \begin{matrix}
        \hat{\psi}_{\uparrow}(x) \\
        \hat{\psi}_{\downarrow}^{\dagger}(x)
    \end{matrix} \right]
    =
    \sum_{\nu}
    \left[ \begin{matrix}
        u_{\nu}^{(0)}(\mathbf{r}) \\
        v_{\nu}^{(0)}(\mathbf{r})
    \end{matrix} \right]
    e^{-iE_{\nu}t/\hbar}
    \hat{A}_{\nu}(0).
\label{operator_steady}
\end{equation}
We can confirm that the Fermi
field operator (\ref{operator_steady})
satisfies the Heisenberg
equation (\ref{Heisenberg})
by using the time-independent BdG equations
(\ref{stedy_BdG}).

The Fermi operator $\hat{A}_{\nu}(0)$ allows
the explicit calculation of the
order parameter and density function
\begin{align}
    \Delta^{(0)}(\mathbf{r})
    &=
    -g \sum_{\nu}
    u_{\nu}^{(0)}(\mathbf{r})
    v_{\nu}^{(0)*}(\mathbf{r})
    f_{\nu},
    \label{Delta_steady} \\
    n^{(0)}(\mathbf{r})
    &=
    \sum_{\nu}
    \left| u_{\nu}^{(0)}(\mathbf{r}) \right|^2
    f_{\nu} \label{n_steady},
\end{align}
where
\begin{equation}
    \Braket{\hat{A}_{\mu}^{\dagger}(0)\hat{A}_{\nu}(0)}
    =
    \delta_{\mu\nu} f_{\nu},
    \quad
    \Braket{\hat{A}_{\mu}(0)\hat{A}_{\nu}^{\dagger}(0)}
    =
    \delta_{\mu\nu} (1-f_{\nu}),
\end{equation}
and $f_{\nu}$ is the Fermi-Dirac distribution
\begin{equation}
    f_{\nu}
    \equiv
    f(E_{\nu})
    =   \frac{1}{1+e^{\beta E_{\nu}}}
    , \quad
    f_{-\nu} = 1 - f_{\nu}.
\end{equation}
The stationary solutions
$\{ u_{\nu}^{(0)}, v_{\nu}^{(0)}, E_{\nu} \}$ can be
obtained by solving the time-independent BdG equations
(\ref{stedy_BdG}), the order parameter
(\ref{Delta_steady}), and the density function (\ref{n_steady})
self-consistently.

\subsection{Perturbations around steady state}
The Fermi field operators $\hat{\psi}_{\sigma}(x)$
are expanded
in terms of $\hat{A}_{\nu}(0)$ as follows:
\begin{equation}
    \left[ \begin{matrix}
        \hat{\psi}_{\uparrow}(x) \\
        \hat{\psi}_{\downarrow}^{\dagger}(x)
    \end{matrix} \right]
    =
    \sum_{\nu}
    \left[ \begin{matrix}
        u_{\nu}(x) \\
        v_{\nu}(x)
    \end{matrix} \right]
    \hat{A}_{\nu}(0),
\end{equation}
where $u_{\nu}(x)$ and $v_{\nu}(x)$ at each time $t$
satisfy the time-dependent BdG equations
\begin{equation}
    i\hbar\frac{\partial}{\partial t}
    \left[ \begin{matrix}
        u_{\nu}(x) \\
        v_{\nu}(x)
    \end{matrix} \right]
    =
    \left[ \begin{matrix}
        \mathcal{L}(x) & \Delta(x) \\
        \Delta^*(x) & -\mathcal{L}(x)
    \end{matrix} \right]
    \left[ \begin{matrix}
        u_{\nu}(x) \\
        v_{\nu}(x)
    \end{matrix} \right],
\label{tdBdG}
\end{equation}
with
\begin{align}
    \Delta(x)
    &=  -g \sum_{\nu} u_{\nu}(x) v_{\nu}^*(x) f_{\nu}, \\
    n(x)
    &=  \sum_{\nu}
        \left|u_{\nu}(x)\right|^2f_{\nu}.
\end{align}
The dynamics of Fermi superfluids is described
by the time-dependent BdG equations
within the Hartree-Fock approximation
\cite{deGennes1966, Challis2007}.

We consider small fluctuations around
stationary solutions
\begin{equation}
    \left[ \begin{matrix}
        u_{\nu}(x) \\
        v_{\nu}(x)
    \end{matrix} \right]
    =
    \left(
    \left[ \begin{matrix}
        u_{\nu}^{(0)}(\mathbf{r}) \\
        v_{\nu}^{(0)}(\mathbf{r})
    \end{matrix} \right]
    +
    \left[ \begin{matrix}
        \delta u_{\nu}(x) \\
        \delta v_{\nu}(x)
    \end{matrix} \right]
    \right)
    e^{-iE_{\nu}t/\hbar},
\end{equation}
\begin{equation}
    \left[ \begin{matrix}
        \Delta(x) \\
        n(x)
    \end{matrix} \right]
    =
    \left[ \begin{matrix}
        \Delta^{(0)}(\mathbf{r}) \\
        n^{(0)}(\mathbf{r})
    \end{matrix} \right]
    +
    \left[ \begin{matrix}
        \delta \Delta(x) \\
        \delta n(x)
    \end{matrix} \right],
\label{fluctuation_OP}
\end{equation}
where
\begin{align}
    \delta\Delta(x)
    &=
    -g \sum_{\nu}
    \left(
        u_{\nu}^{(0)}(\mathbf{r}) \delta v_{\nu}^*(x)
        +
        v_{\nu}^{(0)*}(\mathbf{r}) \delta u_{\nu}(x)
    \right)
    f_{\nu}, \\
    \delta n(x)
    &=
    \sum_{\nu}
    \left(
        u_{\nu}^{(0)}(\mathbf{r}) \delta u_{\nu}^*(x)
        +
        u_{\nu}^{(0)*}(\mathbf{r}) \delta u_{\nu}(x)
    \right)
    f_{\nu}.
\end{align}
Here, the symbol $\delta$ represents a small deviation.
Expanding the time-dependent
BdG equations (\ref{tdBdG}) in order of $\delta$,
we obtain the time-independent
BdG equations (\ref{stedy_BdG})
from the zeroth order and
linearized equations for the fluctuations
\begin{equation}
\begin{split}
    &\left(
        i\hbar\frac{\partial}{\partial t}
        +
        E_{\nu}
    \right)
    \left[ \begin{matrix}
        \delta u_{\nu}(x) \\
        \delta v_{\nu}(x)
    \end{matrix} \right] \\
    &=
    \left[ \begin{matrix}
        \mathcal{L}^{(0)}(\mathbf{r}) &
        \Delta^{(0)}(\mathbf{r}) \\
        \Delta^{(0)*}(\mathbf{r}) &
        -\mathcal{L}^{(0)}(\mathbf{r})
    \end{matrix} \right]
    \left[ \begin{matrix}
        \delta u_{\nu}(x) \\
        \delta v_{\nu}(x)
    \end{matrix} \right] \\
    &\quad+
    \left[ \begin{matrix}
        -g\delta n(x) & \delta\Delta(x) \\
        \delta\Delta^*(x) & g\delta n(x)
    \end{matrix} \right]
    \left[ \begin{matrix}
        u_{\nu}^{(0)}(\mathbf{r}) \\
        v_{\nu}^{(0)}(\mathbf{r})
    \end{matrix} \right],
\label{linear_BdG}
\end{split}
\end{equation}
up to the first order of $\delta$.
We express the order parameter in polar form
and introduce amplitude and phase fluctuations as
\begin{equation}
\begin{split}
    &\Delta(x) \\
    &=  |\Delta(x)|e^{i\theta(x)} \\
    &=  \left(
            \rho^{(0)}(\mathbf{r}) + \delta\rho(x)
        \right)
        e^{i(
            \theta^{(0)}(\mathbf{r})
            +
            \delta\theta(x)/\rho^{(0)}(\mathbf{r})
        )} \\
    &=  \rho^{(0)}(\mathbf{r}) e^{i\theta^{(0)}(\mathbf{r})}
        +
        e^{i\theta^{(0)}(\mathbf{r})}
        \Big(
            \delta\rho(x)
            +
            i\delta\theta(x)
        \Big)
        +
        \mathcal{O}(\delta^2).
\end{split}
\label{fluctuation_OP_rad}
\end{equation}
Here, $\rho$ and $\theta$ represent
the amplitude and phase, respectively.
Equation (\ref{fluctuation_OP}) and (\ref{fluctuation_OP_rad}) imply
\begin{equation}
    \Delta^{(0)}(\mathbf{r})
    =   \rho^{(0)}(\mathbf{r})
        e^{i\theta^{(0)}(\mathbf{r})},
\end{equation}
\begin{equation}
    \delta\Delta(x)
    =
    e^{i\theta^{(0)}(\mathbf{r})}
    \Big(
        \delta\rho(x)
        +
        i\delta\theta(x)
    \Big).
\end{equation}
Expanding $[\delta u_{\nu} \,\, \delta v_{\nu}]^T$
in terms of the eigenfunctions of
the time-independent BdG equations
\begin{equation}
    \left[ \begin{matrix}
        \delta u_{\nu}(x) \\
        \delta v_{\nu}(x)
    \end{matrix} \right]
    =
    \sum_{\rho}
    c_{\nu\rho}(t)
    \left[ \begin{matrix}
        u_{\rho}^{(0)}(\mathbf{r}) \\
        v_{\rho}^{(0)}(\mathbf{r})
    \end{matrix} \right],
\label{complete}
\end{equation}
we obtain
\begin{equation}
\begin{split}
    &\left(
        i\hbar\frac{\partial}{\partial t}
        +
        E_{\nu}
        -
        E_{\rho}
    \right)
    c_{\nu\rho}(t) \\
    &=
    \int d^3\mathbf{r}
    \left(
        W_{\rho\nu}^*(\mathbf{r})
        +
        W_{\nu\rho}(\mathbf{r})
    \right)
    \delta\rho(x) \\
    &+
    \int d^3\mathbf{r}
    \left(
        W_{\rho\nu}^*(\mathbf{r})
        -
        W_{\nu\rho}(\mathbf{r})
    \right)
    i\delta\theta(x) \\
    &+
    \int d^3\mathbf{r} \,
    T_{\nu\rho}(\mathbf{r})
    \big( -g\delta n(x) \Big),
\end{split}
\label{diff_eq_c}
\end{equation}
\begin{equation}
\begin{split}
    \left[ \begin{matrix}
        \delta\rho(x) \\
        i\delta\theta(x) \\
        -gn(x)
    \end{matrix} \right]
    &=  -\frac{g}{2} \sum_{\nu, \rho}
        \Big(
            f_{\nu} c_{\nu\rho}(t)
            +
            f_{\rho} c_{\rho\nu}^*(t)
        \Big) \\
        &\quad\times
        \left[ \begin{matrix}
            W_{\rho\nu}(\mathbf{r})
            +
            W_{\nu\rho}^*(\mathbf{r}) \\
            W_{\rho\nu}(\mathbf{r})
            -
            W_{\nu\rho}^*(\mathbf{r}) \\
            U_{\nu\rho}(\mathbf{r})
        \end{matrix} \right],
\end{split}
\label{fluc_relation}
\end{equation}
where
\begin{align}
    W_{\nu\rho}(\mathbf{r})
    &=
    u_{\nu}^{(0)}(\mathbf{r}) v_{\rho}^{(0)*}(\mathbf{r})
    e^{-i\theta^{(0)}(\mathbf{r})}, \\
    U_{\nu\rho}(\mathbf{r})
    &=
    2u_{\nu}^{(0)*}(\mathbf{r}) u_{\rho}^{(0)}(\mathbf{r}), \\
    T_{\nu\rho}(\mathbf{r})
    &=
    u_{\nu}^{(0)}(\mathbf{r}) u_{\rho}^{(0)* }(\mathbf{r})
    -
    v_{\nu}^{(0)}(\mathbf{r}) v_{\rho}^{(0)* }(\mathbf{r}).
\end{align}
We focus on the Higgs and NG modes through the
two-mode approximation
\cite{Dutta2017}
\begin{equation}
    \left[ \begin{matrix}
        \delta\rho(x) \\
        \delta\theta(x) \\
        \delta n(x) \\
        c_{\nu\rho}(t)
    \end{matrix} \right]
    \approx
    \left[ \begin{matrix}
        \delta\rho(\mathbf{r}) \\
        \delta\theta(\mathbf{r}) \\
        \delta n(\mathbf{r}) \\
        c_{\nu\rho}^{(+)}
    \end{matrix} \right]
    e^{-i\omega t}
    +
    \left[ \begin{matrix}
        \delta\rho^*(\mathbf{r}) \\
        \delta\theta^*(\mathbf{r}) \\
        \delta n^*(\mathbf{r}) \\
        c_{\nu\rho}^{(-)*}
    \end{matrix} \right]
    e^{i\omega t},
\label{two-mode_approx}
\end{equation}
where $\omega$ is real.
The two-mode approximation
describes a system using only two dominant
modes to keep essential physics
while simplifying the analysis.
Substituting Eq.~(\ref{two-mode_approx})
into Eq.~(\ref{diff_eq_c}) and~(\ref{fluc_relation}),
we obtain
\begin{equation}
\begin{split}
    &\Big( E_{\nu} - E_{\rho} + \hbar\omega \Big)
    c_{\nu\rho}^{(+)} \\
    &=  -\Big( E_{\nu} - E_{\rho} + \hbar\omega \Big)
        c_{\rho\nu}^{(-)*} \\
    &=  \int d^3\mathbf{r}
        \left[ \begin{matrix}
            \big\{
                \mathcal{P}_{\nu\rho}^1(\mathbf{r})
            \big\}^* &
            \big\{
                \mathcal{P}_{\nu\rho}^2(\mathbf{r})
            \big\}^* &
            T_{\nu\rho}(\mathbf{r})
        \end{matrix} \right]
        \left[ \begin{matrix}
            \delta\rho(\mathbf{r}) \\
            i\delta\theta(\mathbf{r}) \\
            -g\delta n(\mathbf{r})
        \end{matrix} \right],
\end{split}
\end{equation}
\begin{equation}
\begin{split}
    \left[ \begin{matrix}
        \delta\rho(\mathbf{r}) \\
        i\delta\theta(\mathbf{r}) \\
        -gn(\mathbf{r})
    \end{matrix} \right]
    &=  -\frac{g}{2} \sum_{\nu, \rho}
        \Big(
            f_{\nu} c_{\nu\rho}^{(+)}
            +
            f_{\rho} c_{\rho\nu}^{(-)*}
        \Big) \\
        &\quad\times
        \left[ \begin{matrix}
            W_{\rho\nu}(\mathbf{r})
            +
            W_{\nu\rho}^*(\mathbf{r}) \\
            W_{\rho\nu}(\mathbf{r})
            -
            W_{\nu\rho}^*(\mathbf{r}) \\
            U_{\nu\rho}(\mathbf{r})
        \end{matrix} \right],
\end{split}
\label{fluc_relation_2}
\end{equation}
where
\begin{equation}
    \mathcal{P}_{\nu\rho}^i(\mathbf{r})
    =
    W_{\rho\nu}(\mathbf{r})
    +
    (-1)^{i+1}
    W_{\nu\rho}^*(\mathbf{r}).
\end{equation}
Thus, the desired integral equations are given by
\begin{equation}
\begin{split}
    \left[ \begin{matrix}
        \delta\rho(\mathbf{r}) \\
        i\delta\theta(\mathbf{r}) \\
        -g\delta n(\mathbf{r})
    \end{matrix} \right]
    &=
    -\frac{g}{2}
    \sum_ {\nu, \rho}
    \int d^3 \mathbf{r}'
    \frac{f_{\nu} - f_{\rho}}
    {E_{\nu} - E_{\rho} + \hbar\omega} \\
    &\quad\times
    \mathbf{P}_{\nu\rho}(\mathbf{r}, \mathbf{r}')
    \left[ \begin{matrix}
        \delta\rho(\mathbf{r}') \\
        i\delta\theta(\mathbf{r}') \\
        -g\delta n(\mathbf{r}')
    \end{matrix} \right],
\end{split}
\label{mode_equation}
\end{equation}
where
\begin{equation}
\begin{split}
    &\mathbf{P}_{\nu\rho}(\mathbf{r}, \mathbf{r}') \\
    &=
    \left[ \begin{matrix}
        \mathcal{P}_{\nu\rho}^1(\mathbf{r}) \\
        \mathcal{P}_{\nu\rho}^2(\mathbf{r}) \\
        U_{\nu\rho}(\mathbf{r})
    \end{matrix} \right]
    \left[ \begin{matrix}
        \big\{
            \mathcal{P}_{\nu\rho}^1(\mathbf{r}')
        \big\}^* &
        \big\{
            \mathcal{P}_{\nu\rho}^2(\mathbf{r}')
        \big\}^* &
        T_{\nu\rho}(\mathbf{r}')
    \end{matrix} \right],
\end{split}
\end{equation}
The integral equations (\ref{mode_equation})
incorporate the spatial dependence of the
trapped Fermi superfluids.
These equations admit nontrivial solutions corresponding
to the Higgs and NG modes when $\delta\rho$ and
$\delta\theta$ are nonzero, respectively.
In the following sections, we solve the integral
equations (\ref{mode_equation}) to obtain these
collective modes.

\section{Solving integral equation}
\label{solve_equation}

This section first presents analytical results
for homogeneous systems and then provides numerical
results for inhomogeneous systems.

\subsection{Homogeneous system}
By applying the integral equations (\ref{mode_equation})
to a homogeneous system,
we confirm the existence of nontrivial solutions
corresponding to the Higgs and NG modes.
By setting $V_{\text{trap}}=\delta n=0$,
we rewrite the integral equations~(\ref{mode_equation})
by performing the Fourier transformation as
\begin{equation}
\begin{split}
    \left[ \begin{matrix}
        \delta\rho(\mathbf{k}) \\
        i\delta\theta(\mathbf{k})
    \end{matrix} \right]
    &=
    -\frac{g}{2V} \sum_{\mathbf{p}, \sigma_+, \sigma_-}
    \frac{f(E_{\sigma_+})-f(E_{\sigma_-})}
    {E_{\sigma_+}-E_{\sigma_-}+\hbar\omega} \\
    &\quad\times
    \left[ \begin{matrix}
        \mathcal{P}_{\sigma_+\sigma_-}^1
        \mathcal{P}_{\sigma_+\sigma_-}^1 &
        \mathcal{P}_{\sigma_+\sigma_-}^1
        \mathcal{P}_{\sigma_+\sigma_-}^2 \\
        \mathcal{P}_{\sigma_+\sigma_-}^2
        \mathcal{P}_{\sigma_+\sigma_-}^1 &
        \mathcal{P}_{\sigma_+\sigma_-}^2
        \mathcal{P}_{\sigma_+\sigma_-}^2
    \end{matrix} \right]
    \left[ \begin{matrix}
        \delta\rho(\mathbf{k}) \\
        i\delta\theta(\mathbf{k})
    \end{matrix} \right],
\end{split}
\end{equation}
where
\begin{equation}
\begin{split}
    \mathcal{P}_{\sigma_+\sigma_-}^i
    &=
    \left\{ \begin{matrix}
        u_{\mathbf{p}_-} \\
        -v_{\mathbf{p}_-}
    \end{matrix} \right\}^{\sigma_-}
    \left\{ \begin{matrix}
        v_{\mathbf{p}_+} \\
        u_{\mathbf{p}_+}
    \end{matrix} \right\}^{\sigma_+} \\
    &\quad+
    (-1)^{i+1}
    \left\{ \begin{matrix}
        u_{\mathbf{p}_+} \\
        -v_{\mathbf{p}_+}
    \end{matrix} \right\}^{\sigma_+}
    \left\{ \begin{matrix}
        v_{\mathbf{p}_-} \\
        u_{\mathbf{p}_-}
    \end{matrix} \right\}^{\sigma_-},
\end{split}
\end{equation}
with
$E_{\sigma_{\pm}}
=
\{
E_{\mathbf{p}_{\pm}}, -E_{\mathbf{p}_{\pm}} 
\}^{\sigma_{\pm}}
$
and
$\mathbf{p}_{\pm}=\mathbf{p}\pm\mathbf{k}/2$.
The BdG functions $\{ u_{\mathbf{p}}, v_{\mathbf{p}} \}$
are given by
\begin{equation}
    u_{\mathbf{p}}
    =   \sqrt{
            \frac{1}{2}
            \left(
                1
                +
                \frac{\xi_{\mathbf{p}}}{E_{\mathbf{p}}}
            \right)
        }
    , \quad
    v_{\mathbf{p}}
    =   \sqrt{
            \frac{1}{2}
            \left(
                1
                -
                \frac{\xi_{\mathbf{p}}}{E_{\mathbf{p}}}
            \right)
        },
\end{equation}
with $\xi_{\mathbf{p}}=\hbar^2\mathbf{p}^2/2m-\mu$ and
$E_{\mathbf{p}}=\sqrt{\xi_{\mathbf{p}}^2+|\Delta^{(0)}|^2}$.
By using the gap equation
\begin{equation}
    1
    =
    \frac{g}{V} \sum_{\mathbf{p}}
    \frac{\tanh(\beta E_{\mathbf{p}}/2)}{2E_{\mathbf{p}}},
\label{gap_equation}
\end{equation}
we obtain the following equation as $\mathbf{k} \to 0$:
\begin{equation}
    \left[ \begin{matrix}
        F(\omega)
        \left\{
            (\hbar\omega)^2 - (2|\Delta^{(0)}|)^2
        \right\} &
        2\hbar\omega G(\omega) \\
        2\hbar\omega G(\omega) &
        F(\omega)(\hbar\omega)^2
    \end{matrix} \right]
    \left[ \begin{matrix}
        \delta\rho(\boldsymbol{0}) \\
        i\delta\theta(\boldsymbol{0}) \\
    \end{matrix} \right]
    =   \boldsymbol{0},
\label{mode_equation_uniform}
\end{equation}
where
\begin{align}
    F(\omega)
    &=
    \frac{g}{V} \sum_{\mathbf{p}}
    \frac{\tanh(\beta E_{\mathbf{p}}/2)}{2E_{\mathbf{p}}}
    \frac{1}{4E_{\mathbf{p}}^2-(\hbar\omega)^2}, \\
    G(\omega)
    &=
    \frac{g}{V} \sum_{\mathbf{p}}
    \frac{\tanh(\beta E_{\mathbf{p}}/2)}{2E_{\mathbf{p}}}
    \frac{\xi_{\mathbf{p}}}
    {4E_{\mathbf{p}}^2-(\hbar\omega)^2}.
\end{align}
The density of states can be approximated
as nearly constant. This approximation arises
from converting the summation into an integral
over $\xi_{\mathbf{p}}$,
and the integral range can be taken
symmetrically around $\xi_{\mathbf{p}}=0$.
Since the integrands of $F$ and $G$ are the
even and odd functions of $\xi_{\mathbf{p}}$,
respectively, we obtain
\begin{equation}
    \left[ \begin{matrix}
        (\hbar\omega)^2 - (2|\Delta^{(0)}|)^2 & 0 \\
        0 & (\hbar\omega)^2
    \end{matrix} \right]
    \left[ \begin{matrix}
        \delta\rho(\boldsymbol{0}) \\
        i\delta\theta(\boldsymbol{0}) \\
    \end{matrix} \right]
    =   \boldsymbol{0},
\label{mode_equation_uniform_approx}
\end{equation}
with $G(\omega) \approx 0$ and $F(\omega) \neq 0$.
Equation (\ref{mode_equation_uniform_approx})
shows that
the Higgs mode $\omega=2|\Delta^{(0)}|/\hbar$
and NG mode $\omega=0$ result from
the amplitude and phase
fluctuations, respectively.
These results are consistent with the previous works
~\cite{Anderson1958, Littlewood1981, Littlewood1982,
Kulik1981, Tsuji2015, Volkov1974}.
We note that
$\omega=0$ is an exact solution of the characteristic equation
for non-vanishing $G(\omega)$
\begin{equation}
    \operatorname{det}
    \left[ \begin{matrix}
        F(\omega)
        \left\{
            (\hbar\omega)^2 - (2|\Delta^{(0)}|)^2
        \right\} &
        2\hbar\omega G(\omega) \\
        2\hbar\omega G(\omega) &
        F(\omega)(\hbar\omega)^2
    \end{matrix} \right]
    =   0.
\end{equation}

\subsection{Inhomogeneous system}
In this section, we solve the integral equations~(\ref{mode_equation})
in an inhomogeneous system through numerical calculations,
with an isotropic harmonic potential as a typical example.
Assuming that the fluctuations depend only on
the radial coordinate in polar coordinates and setting $\delta n=0$,
we obtain
\begin{equation}
\begin{split}
    &\left[ \begin{matrix}
        \delta\rho(r) \\
        i\delta\theta(r)
    \end{matrix} \right]
    =
    -\frac{g}{2}
    \sum_{nn'l} \frac{2l+1}{4\pi}
    \int dr' {r'}^2
    \frac{f(E_{nl}) - f(E_{n'l})}
    {E_{nl} - E_{n'l} + \hbar\omega} \\
    &\times
    \left[ \begin{matrix}
        \mathcal{P}_ {nn'l}^1(r) \\
        \mathcal{P}_ {nn'l}^2(r)
    \end{matrix} \right]
    \left[ \begin{matrix}
        \big\{ \mathcal{P}_ {nn'l}^1(r') \big\}^* &
        \big\{ \mathcal{P}_ {nn'l}^2(r') \big\}^*
    \end{matrix} \right]
    \left[ \begin{matrix}
        \delta\rho(r') \\
        i\delta\theta(r')
    \end{matrix} \right],
\end{split}
\label{mode_equation_radial}
\end{equation}
where
\begin{equation}
\begin{split}
    \mathcal{P}_ {nn'l}^i(r)
    &=  u_{n'l}(r) v_{nl}^* (r) e^{-i\theta(r)} \\
        &\quad+
        (-1)^{i+1}
        u_{nl}^* (r) v_{n'l}(r) e^{i\theta(r)}.
\end{split}
\end{equation}
Here, $\{ u_{nl}, v_{nl}, E_{nl} \}$
satisfies
\begin{equation}
    \left[ \begin{matrix}
        \mathcal{L}_l(r) &
        \Delta(r) \\
        \Delta^*(r) &
        -\mathcal{L}_l(r)
    \end{matrix} \right]
    \left[ \begin{matrix}
        u_{nl}(r) \\
        v_{nl}(r)
    \end{matrix} \right]
    =
    E_{nl}
    \left[ \begin{matrix}
        u_{nl}(r) \\
        v_{nl}(r)
    \end{matrix} \right],
\label{radial_BdG}
\end{equation}
with
\begin{align}
    \mathcal{L}_l(r)
    &=  -\frac{\hbar^2}{2m}
        \left(
            \frac{d^2}{d r^2}
            +
            \frac{2}{r} \frac{d}{d r}
            -
            \frac{l(l+1)}{r^2}
        \right) \notag \\
        &\quad+
        \frac{m\omega_{\text{ho}}^2r^2}{2}
        -
        \mu
        -
        gn(r), \\
    \Delta(r)
    &=  -g \sum_{nl} \frac{2l+1}{4\pi}
        u_{nl}(r) v_{nl}^*(r) f(E_{nl}), \\
    n(r)
    &=  \sum_{nl} \frac{2l+1}{4\pi}
        |u_{nl}(r)|^2 f(E_{nl}).
\end{align}
The integers $n$ and $l$ are the radial and
azimuthal quantum numbers, respectively.
The frequency $\omega_{\text{ho}}$
characterizes the strength of the trapping potential.

To perform numerical calculations,
we discretize the integral equation
(\ref{mode_equation_radial}) as
\begin{equation}
\begin{split}
    &\left[ \begin{matrix}
        \delta\rho(r_i) \\
        i\delta\theta(r_i)
    \end{matrix} \right] \\
    &=
    \sum_{j=1}^M
    \left[ \begin{matrix}
        \mathcal{M}_{11}(r_i, r_j; \omega) &
        \mathcal{M}_{12}(r_i, r_j; \omega) \\
        \mathcal{M}_{21}(r_i, r_j; \omega) &
        \mathcal{M}_{22}(r_i, r_j; \omega)
    \end{matrix} \right]
    \left[ \begin{matrix}
        \delta\rho(r_j) \\
        i\delta\theta(r_j)
    \end{matrix} \right],
\end{split}
\end{equation}
where 
\begin{equation}
\begin{split}
    \mathcal{M}_{ij}(r, r'; \omega)
    &=
    -\frac{g}{2}
    \sum_{nn'l} \frac{2l+1}{4\pi}
    \frac{f(E_{nl}) - f(E_{n'l})}
    {E_{nl} - E_{n'l} + \hbar\omega} \\
    &\quad\times
    \mathcal{P}_ {nn'l}^i(r)
    \big\{ \mathcal{P}_ {nn'l}^j(r') \big\}^*
    {r'}^2 \Delta r.
\end{split}
\end{equation}
Here, $M$ and $\Delta r$ are the number of spatial
divisions and resolution, respectively.
For simplicity,
we assume that the off-diagonal elements
$\mathcal{M}_{12}$ and $\mathcal{M}_{21}$
are small enough to be approximated as zero,
as in the case of the uniform system.
To determine the frequency $\omega$ of the Higgs mode
by numerical calculation,
we solve the following characteristic equation
\begin{equation}
    \operatorname{det}
    \Big[
        \mathbf{I}
        -
        \mathbf{M}_{11}(\omega)
    \Big]
    =   0,
\end{equation}
where $\mathbf{I}$ is the identity matrix and
\begin{equation}
    \mathbf{M}_{ij}(\omega)
    =
    \left[ \begin{matrix}
        \mathcal{M}_{ij}(r_1, r_1; \omega) &
        \cdots &
        \mathcal{M}_{ij}(r_1, r_M; \omega) \\
        \vdots &
        \ddots &
        \vdots \\
        \mathcal{M}_{ij}(r_M, r_1; \omega) &
        \cdots &
        \mathcal{M}_{ij}(r_M, r_M; \omega)
    \end{matrix} \right].
\end{equation}

An isotropic harmonic trapped system is characterized
by the scattering length $a_S$, the frequency of the
trapping potential $\omega_{\text{ho}}$, and
the temperature $T$.
The coupling constant $g$ is calculated by the scattering length $a_S$.
For numerical calculations, the three independent parameters are
expressed in terms of three dimensionless parameters
$1/a_Sk_F, 1/a_{\text{ho}}k_F$ and $T/T_F$.
Here,
$a_{\text{ho}}=\sqrt{\hbar/(m\omega_{\text{ho}})}$,
$k_F$, and $T_F$ are
the harmonic oscillator length,
Fermi wave number, and Fermi temperature, respectively.
The Fermi wave number $k_F$ and
Fermi temperature $T_F$ are determined from the
Fermi energy $E_F$.

When calculating the time-independent
BdG equations (\ref{radial_BdG})
and integral equations
(\ref{mode_equation_radial})
numerically, all the sums in these equations
are limited by an energy cutoff $E_c > |E_{nl}|$.
This cutoff is necessary to prevent
the ultraviolet divergences, and regularizes
the coupling constant $g$ according to
Ref. \cite{Antezza2007}
\begin{equation}
    \frac{1}{g}
    =   -\frac{m}{4\pi\hbar^2} \frac{1}{a_S}
        +
        \frac{1}{\sqrt{2}\pi^2}
        \left( \frac{m}{\hbar^2} \right)^{3/2}
        \sqrt{E_c},
\label{regularization_g}
\end{equation}
as explained in Appendix \ref{appendix_cutoff}.

In this paper, all the numerical calculations below are performed
under the fixed total number of perticles $N=20000$,
which is provided by
\begin{equation}
    N
    =   2\int d\mathbf{r} \, n(\mathbf{r}).
\label{particle_number}
\end{equation}
Here, the prefactor 2 accounts for spin degeneracy.
Specifically,
we determine the chemical potential $\mu$ using the bisection method,
setting the convergence condition such that the difference
between the particle number obtained from Eq. (\ref{particle_number})
and the target value is less than $10^{-3}$
of the target value.

\subsubsection{Trap dependence}
\begin{figure}
    \includegraphics[width=\linewidth]
    {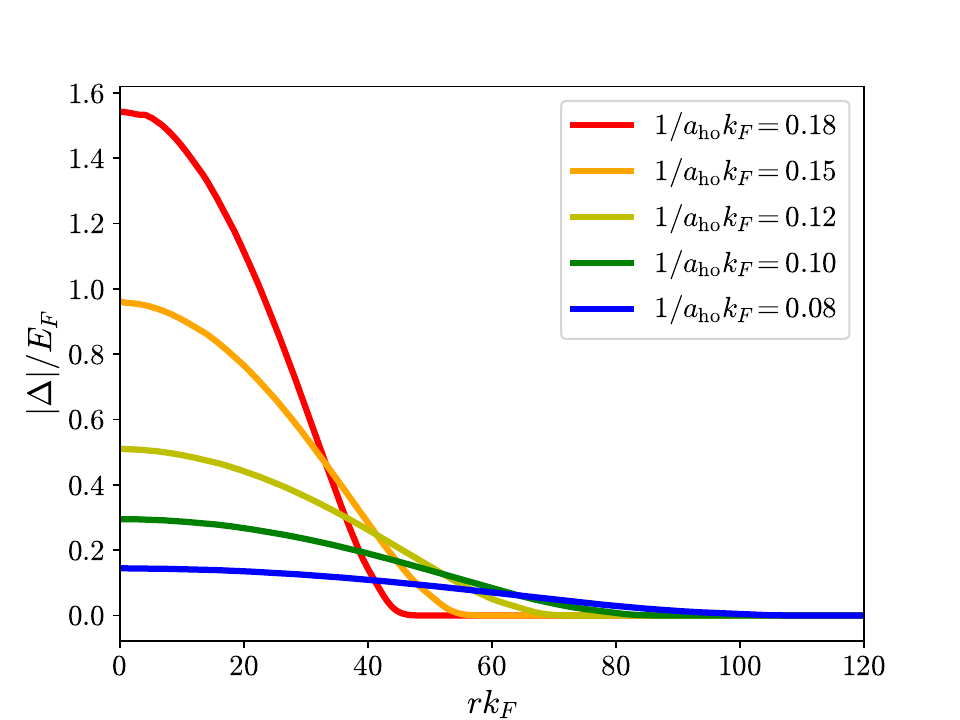}
    % Here is how to import EPS art
    \caption{
        The trap frequency dependence of the order
        parameter. The magnitudes of the order
        parameter are plotted for
        $1/(a_Sk_F)=-0.6$
        and $T/T_F=0$
        as a function of $1/a_{\text{ho}}k_F$:
        red for $1/a_{\text{ho}}k_F=0.18$
        orange for $1/a_{\text{ho}}k_F=0.15$,
        yellow for $1/a_{\text{ho}}k_F=0.12$,
        green for $1/a_{\text{ho}}k_F=0.10$,
        and blue for $1/a_{\text{ho}}k_F=0.08$.
    }
\label{Delta_Odep}
\end{figure}

To confirm the behavior of the Higgs mode when
the trap frequency $\omega_{\text{ho}}$ varies,
we calculate the order parameter and the Higgs mode
frequency, while keeping the scattering length
$a_S$ and temperature $T$ fixed.
Figure \ref{Delta_Odep} shows that
the amplitude of the order parameter $|\Delta|$
becomes flatter as $1/a_{\text{ho}}k_F$ decreases.
This is because $1/a_{\text{ho}}k_F$ is
proportional to $\sqrt{\omega_{\text{ho}}}$, and
the system approaches the uniform limit
as the trap frequency $\omega_{\text{ho}}$ decreases.
On the other hand, the particles become more concentrated
near the center of the trap as $1/a_{\text{ho}}k_F$
increases, resulting in an increase in the central
value of the order parameter.

\begin{figure}
    \includegraphics[width=\linewidth]
    {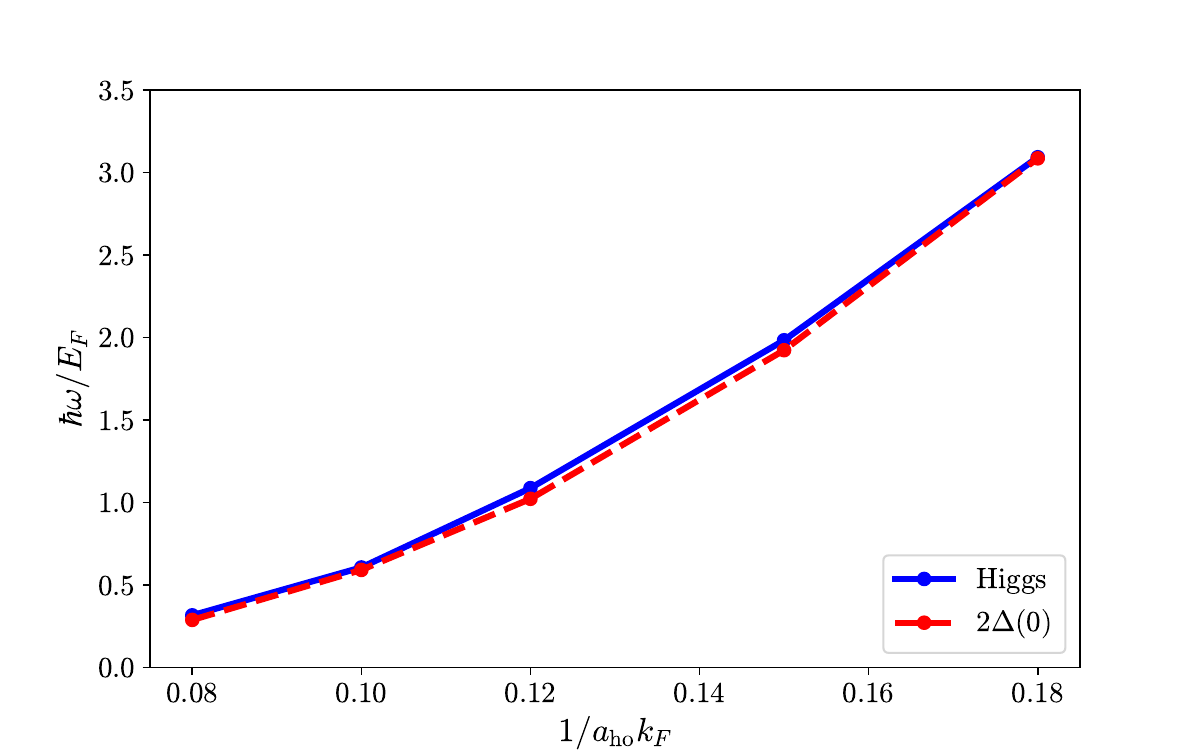}
    % Here is how to import EPS art
    \caption{
        The trap frequency dependence of
        the Higgs mode.
        The frequencies of the Higgs mode in
        the isotropic harmonic trap are plotted for
        $1/(a_Sk_F)=-0.6$
        and $T/T_F=0$
        as a function of $1/a_{\text{ho}}k_F$.
        The blue and red line are
        the frequency of the Higgs mode and
        twice the central value of the order parameter,
        respectively.
    }
\label{Higgs_wdep}
\end{figure}

Figure~\ref{Higgs_wdep} presents the frequency of
the Higgs mode $\omega$ as a monotonically increasing
function of $1/a_{\text{ho}}k_F$.
The key finding is that $\omega = 2|\Delta(0)|$
always holds regardless of the value of $1/a_{\text{ho}}k_F$.
This result is consistent with the previous works
~\cite{Scott2012, Tokimoto2017, Tokimoto2019, Hannibal2018, Bjerlin2016, Bruun2014, Dutta2017}.

\subsubsection{Interaction dependence}
\begin{figure}
    \includegraphics[width=\linewidth]
    {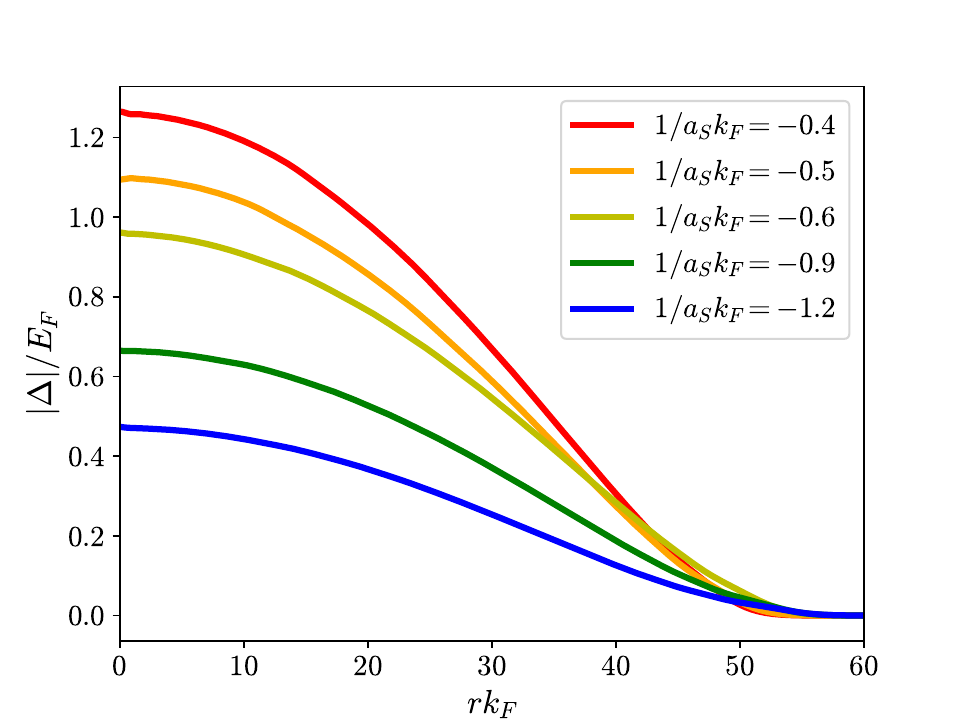}
    % Here is how to import EPS art
    \caption{
        The interaction strength dependence of the order
        parameter. The magnitudes of the order
        parameter are plotted for
        $1/(a_{\text{ho}}k_F)=0.15$
        and $T/T_F=0$
        as a function of $1/a_Sk_F$:
        red for $1/a_Sk_F=-0.4$,
        orange for $1/a_Sk_F=-0.5$,
        yellow for $1/a_Sk_F=-0.6$,
        green for $1/a_Sk_F=-0.9$,
        and blue for $1/a_Sk_F=-1.2$.
    }
\label{Delta_Udep}
\end{figure}

We investigate the behavior of the Higgs mode by
varying only the scattering length $a_S$,
while keeping the trap frequency $\omega_{\text{ho}}$
and temperature $T$ fixed.
Figure~\ref{Delta_Udep} shows that the
overall amplitude of the order parameter $|\Delta|$
increases as $1/a_Sk_F$ increases due to the enhancement
of the interaction strength $g$.
Compared with variations in $1/a_{\text{ho}}k_F$,
the spatial extent of the order parameter remains constant
when $1/a_Sk_F$ is varied.
The behavior of the order parameter
can be attributed to the fact that,
in regions with high particle density near the trap center,
the stronger attractive interaction
facilitates the formation of more Cooper pairs.
In contrast, in regions farther from the center
where the particle density is lower,
the dependence of the strength of the interaction is not remarkable.

\begin{figure}
    \includegraphics[width=\linewidth]
    {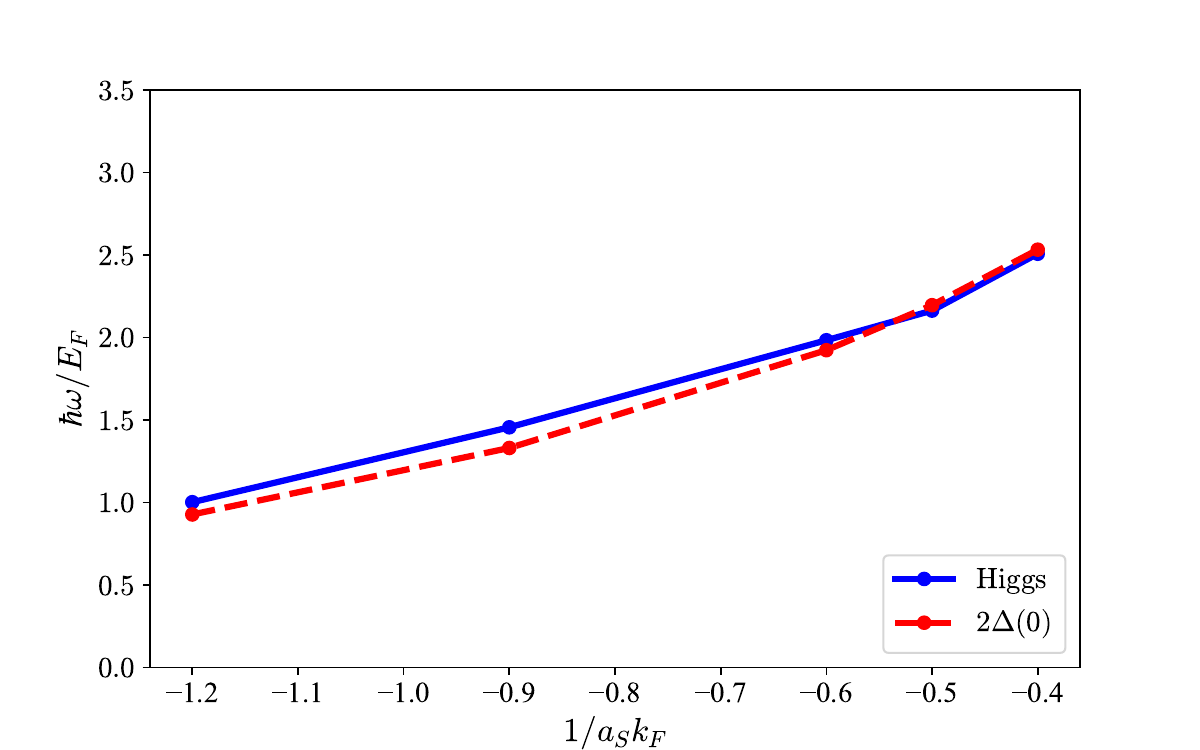}
    % Here is how to import EPS art
    \caption{
        The interaction strength dependence of
        the Higgs mode.
        The frequencies of the Higgs mode in
        the isotropic harmonic trap are plotted for
        $1/(a_{\text{ho}}k_F)=0.15$
        and $T/T_F=0$
        as a function of $1/a_Sk_F$.
        The blue and red line are
        the frequency of the Higgs mode and
        twice the central value of the order parameter,
        respectively.
    }
\label{Higgs_Udep}
\end{figure}

Figure \ref{Higgs_Udep} presents the frequencies of
the Higgs mode for various $1/a_Sk_F$.
Similarly to the case where the trap frequency is varied,
the frequency of the Higgs mode $\omega$ holds to be twice
as large as $|\Delta(0)|$ at all times.
In addition, the frequency of the Higgs mode increases
as $1/a_Sk_F$ increases.

\subsubsection{Temperature dependence}
\begin{figure}
    \includegraphics[width=\linewidth]
    {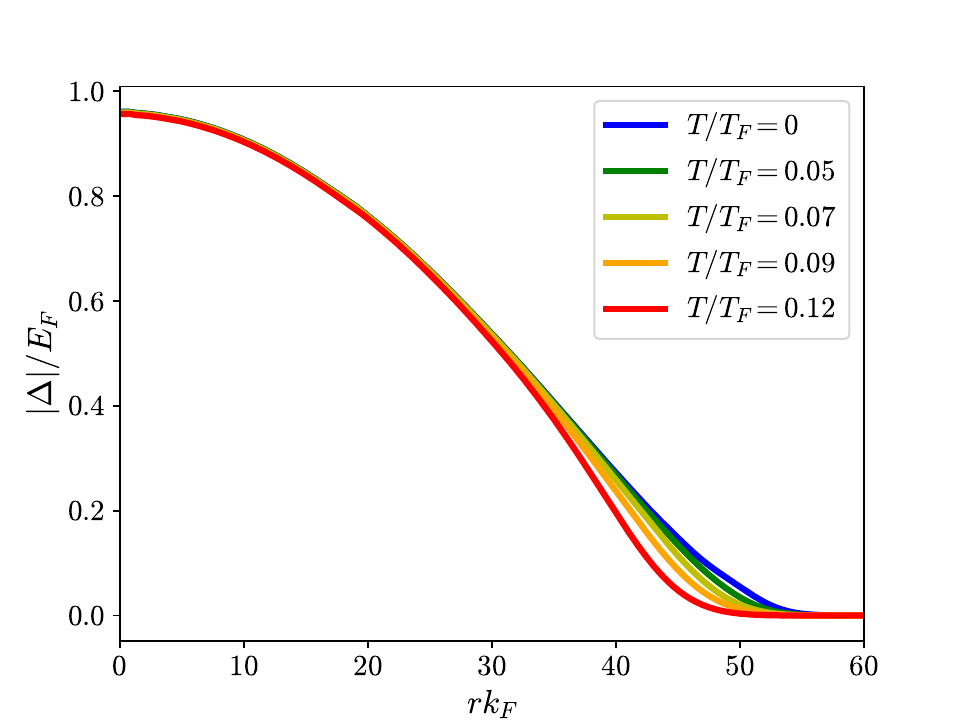}
    % Here is how to import EPS art
    \caption{
        The temperature dependence of the order
        parameter. The magnitudes of the order
        parameter are plotted for
        $1/(a_Sk_F)=-0.6$ and
        $1/(a_{\text{ho}}k_F)=0.15$
        at various temperatures:
        blue for $T/T_F=0$, green for $T/T_F=0.05$,
        yellow for $T/T_F=0.07$, orange for $T/T_F=0.09$,
        and red for $T/T_F=0.12$.
    }
\label{Delta_Tdep}
\end{figure}

The temperature dependence of the order parameter
is shown in Fig. \ref{Delta_Tdep}.
We plot the spatial profile of the order parameter
as a function of temperature $T$, while keeping
the scattering length $a_S$ and the trap frequency
$\omega_{\text{ho}}$ fixed.
As the temperature increases,
the spatial extent of the order parameter shrinks,
while its central value of the order parameter
remains nearly constant.

\begin{figure}
    \includegraphics[width=\linewidth]
    {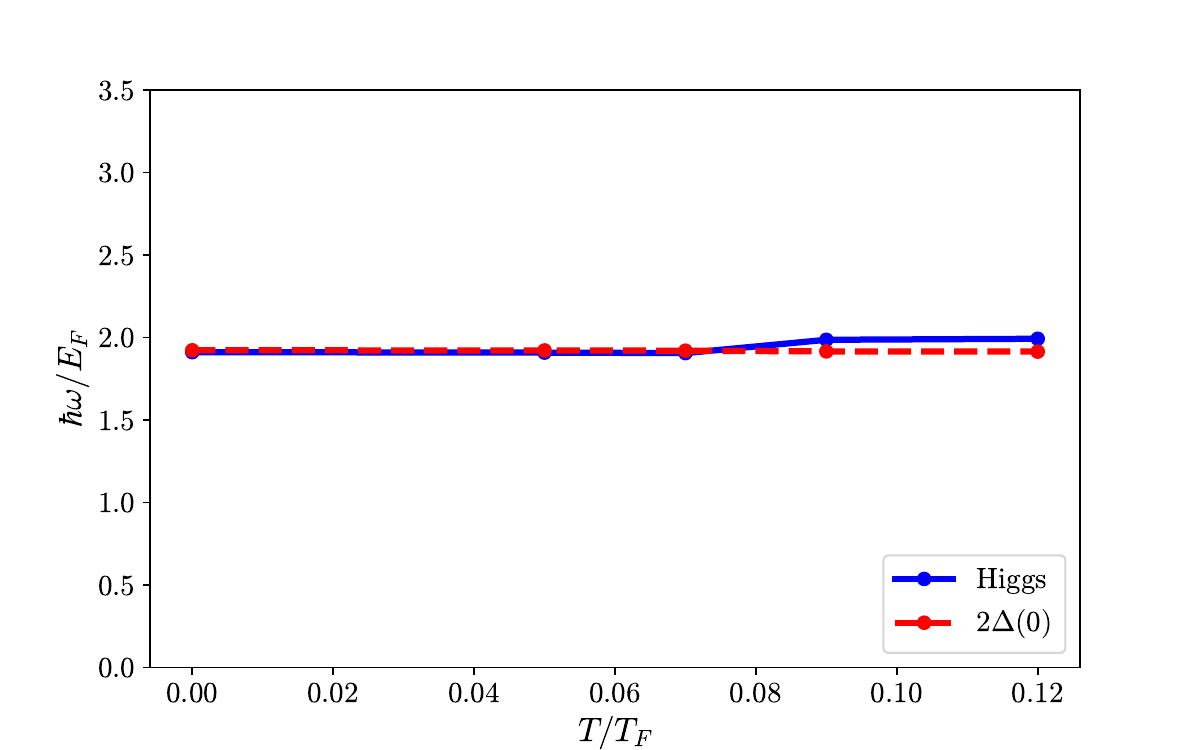}
    % Here is how to import EPS art
    \caption{
        The temperature dependence of the Higgs mode.
        The values of the Higgs mode in
        the isotropic harmonic trap are plotted for
        $1/(a_Sk_F)=-0.6$ and
        $1/(a_{\text{ho}}k_F)=0.15$
        as a function of temperature.
        The blue and red line are
        the frequency of the Higgs mode and
        twice the central value of the order parameter,
        respectively.
    }
\label{Higgs_Tdep}
\end{figure}

Figure \ref{Higgs_Tdep} shows the
temperature dependence of the Higgs mode frequencies.
In the low temperature regime, the frequency
of the Higgs mode is approximately constant
because $\Delta(0)$ remains unchanged and
the frequency of the Higgs mode coincides
with $2|\Delta(0)|$.
This finding agrees with the recent experimental
observation~\cite{Dyke2024, Kell2024}.

\section{Conclusions}
\label{conclusions}
We have derived integral equations that describe
Higgs and NG modes, and fully account for
the effect of external potential on Fermi
superfluids.
By applying these equations to homogeneous systems,
we analytically reproduce the Higgs and NG modes.
In inhomogeneous systems, the
Higgs mode emerges as twice the central value
of the order parameter through numerical calculations.
This feature always holds when each of the parameters
$\omega_{\text{ho}}, a_S$ and $T$ varies, respectively.
In addition, the frequency of the Higgs mode increases
as $\omega_{\text{ho}}$ or $a_S$ increase.
On the other hand,
the Higgs mode is temperature-independent in the low-temperature regime.

Further insights can be extracted from our theoretical framework,
such as the collective mode of density fluctuation $\delta n$, and
the coupling between the Higgs and NG modes mediated by
the off-diagonal elements $\mathcal{M}_{12}$ and $\mathcal{M}_{21}$.
In the presence of relativistic Lorentz symmetry,
the Higgs and NG modes are
decoupled~\cite{Pekker2015},
whereas the Higgs and NG modes are mixed with each other
in cases of no Lorentz symmetry.
The analysis of these collective modes remains an
important direction for future work.

\begin{acknowledgments}
We are also grateful to Prof. T. Yamamoto, Prof. S. Uchino,
Dr. M. Okumura and Y. Kurokawa for discussions.
\end{acknowledgments}

\appendix

\section{Cutoff dependence of order parameter}
\label{appendix_cutoff}
A cutoff is introduced to deal with the infinite sums
in numerical calculations, and the sums are truncated.
We consider the Schr\"{o}dinger equation
describing the relative motion of
two Fermi particles
\begin{equation}
    \left(
        \hat{H}_0 + \hat{V}
    \right)
    \ket{\psi}
    =   E \ket{\psi},
\end{equation}
where
\begin{equation}
    \hat{H}_0
    =   \frac{\hat{\mathbf{p}}^2}{m}
    , \quad
    \hat{V} = -g\delta(\hat{\mathbf{x}}),
\end{equation}
with the momentum operator $\hat{\mathbf{p}}$ and
position operator $\hat{\mathbf{x}}$, respectively.
Using an incident wave $\ket{\phi}$ with
the same energy as the state $\ket{\psi}$,
we obtain the Lippmann-Schwinger equation
\begin{equation}
    \ket{\psi}
    =   \ket{\phi}
        +
        \frac{1}{E-H_0} \hat{V}
        \ket{\psi}.
\label{LS_equation}
\end{equation}
Equation (\ref{LS_equation}) can be rewritten
by $T$-matrix
\begin{equation}
    \hat{T}
    =   \hat{V}
        +
        \hat{V}
        \frac{1}{E-H_0} \hat{T},
\label{LS_equation_T}
\end{equation}
where $T$-matrix is defined
\begin{equation}
    \hat{T} \ket{\phi}
    \equiv
    \hat{V} \ket{\psi}.
\end{equation}

The $T$-matrix provides the expression for
the scattering amplitude
\begin{equation}
    f(\mathbf{k}', \mathbf{k})
    \equiv
    -\frac{m}{4\pi\hbar^2}
    \braket{\mathbf{k}'|\hat{V}|\psi}
    =
    -\frac{m}{4\pi\hbar^2}
    \braket{\mathbf{k}'|\hat{T}|\mathbf{k}},
\end{equation}
where $\ket{\phi}=\ket{\mathbf{k}}$ and
we obtain
$E=\hbar^2\mathbf{k}^2/m
\equiv 2\varepsilon_{\mathbf{k}}$.
In the long-wavelength limit,
the scattering length is defined
\begin{equation}
    a_S
    \equiv
    -
    \lim_{|\mathbf{k}|=|\mathbf{k}'| \to 0}
    f(\mathbf{k}', \mathbf{k}).
\end{equation}
The relation between the scattering length
and $T$-matrix is given by
\begin{equation}
    \frac{4\pi\hbar^2}{m} a_S
    =
    \braket{\mathbf{0}|\hat{T}|\mathbf{0}}.
\label{relation_aS_T}
\end{equation}
Using the Lippmann-Schwinger equation
(\ref{LS_equation_T}), we can extend the
right-hand side of Eq. (\ref{relation_aS_T})
\begin{equation}
\begin{split}
    \frac{4\pi\hbar^2}{m} a_S
    &=
    -g
    +
    (-g) \frac{1}{V} \sum_{\mathbf{k}}^{\omega_c}
    \frac{1}{0-2\varepsilon_{\mathbf{k}}} (-g) \\
    &\quad+
    (-g)
    \left(
        \frac{1}{V} \sum_{\mathbf{k}}^{\omega_c}
        \frac{1}{0-2\varepsilon_{\mathbf{k}}}
        (-g)
    \right)^2
    +
    \cdots,
\end{split}
\end{equation}
where $\omega_c$ is the energy cutoff as the limit
of applicability of the mathematical model,
i.e., the Debye frequency of a superconductor.
The formula for the infinite series
provides
\begin{equation}
    \frac{4\pi\hbar^2}{m} a_S
    =
    \frac{-g}{
        1
        -
        \frac{g}{V}\sum_{\mathbf{k}}^{\omega_c}
        \frac{1}{2\varepsilon_{\mathbf{k}}}
    }.
\label{RG_aS}
\end{equation}

Equation (\ref{RG_aS}) is commonly used for
renormalization.
Replacement of the theoretical parameter $g$ with
the experimental measurement $a_S$ prevents
the divergence of the infinite series.
The right-hand side of the gap equation
(\ref{gap_equation})
does not converge in the limit as
$\omega_c \to \infty$. Using Eq. (\ref{RG_aS}),
we obtain
\begin{equation}
    1
    =   -\frac{4\pi\hbar^2}{m} \frac{a_S}{V}
        \sum_{\mathbf{k}}^{\omega_c}
        \left(
            \frac{\tanh(\beta E_{\mathbf{k}}/2)}{2E_{\mathbf{k}}}
            -
            \frac{1}{2\varepsilon_{\mathbf{k}}}
        \right).
\label{gap_equation_RG}
\end{equation}
The right-hand side of Eq. (\ref{gap_equation_RG})
converges in the limit as $\omega_c \to \infty$,
and we can theoretically remove the energy limitation
$\omega_c$.

\begin{figure}
    \includegraphics[width=\linewidth]
    {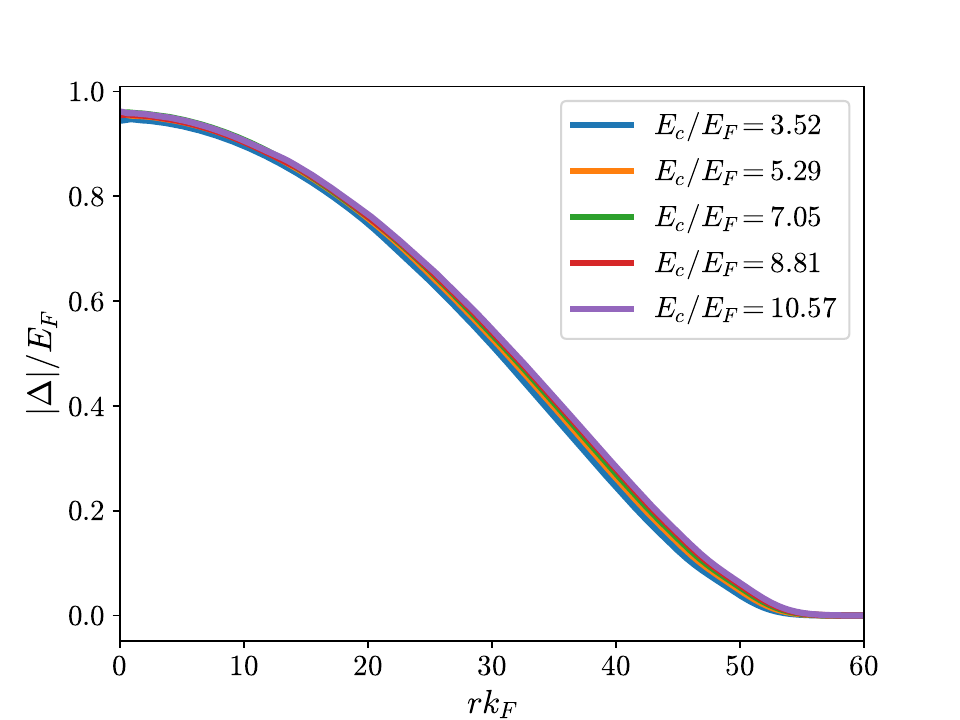}
    % Here is how to import EPS art
    \caption{
        The energy cutoff dependence of the order
        parameter. The magnitudes of the order
        parameter are plotted for
        $1/(a_Sk_F)=-0.6,
        1/(a_{\text{ho}}k_F)=0.15,
        T/T_F=0$ and different energy
        cutoffs $E_c$:
        blue for $E_c/E_F=3.52$
        orange for $E_c/E_F=5.29$,
        green for $E_c/E_F=7.05$,
        red for $E_c/E_F=8.81$,
        and purple for $E_c/E_F=10.57$.
    }
\label{Delta_Ecdep}
\end{figure}

On the other hand, in this paper,
we regard the theoretical cutoff $\omega_c$
as the numerically necessary cutoff $E_c$
and provide the coupling constant regularization
(\ref{regularization_g}).
Figure \ref{Delta_Ecdep} shows the numerical
results of the order parameter with the regularization
(\ref{regularization_g}).
This regularization removes the cutoff dependence
of the order parameter.

\begin{figure}
    \includegraphics[width=\linewidth]
    {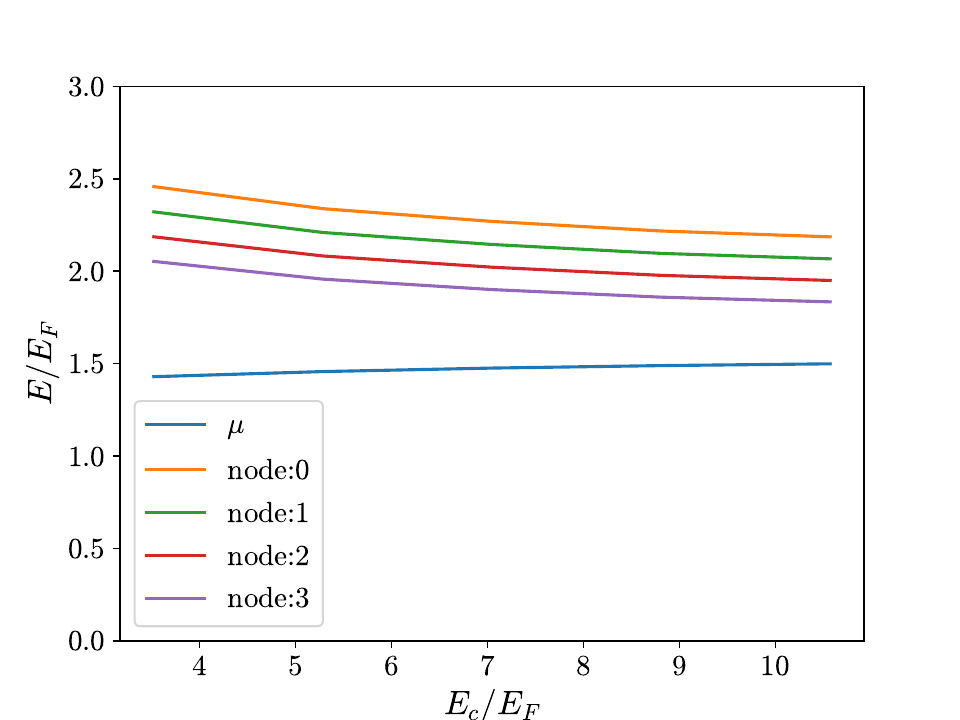}
    % Here is how to import EPS art
    \caption{
        The energy cutoff dependence of
        chemical potential and eigenvalues for
        eigenfunction with 0 to 3 nodes at $l=0$.
        The chemical potential and the energy spectrum
        are plotted for
        $1/(a_Sk_F)=-0.6,
        1/(a_{\text{ho}}k_F)=0.15,
        T/T_F=0$ as a function of $E_c$.
    }
\label{mu_Ecdep}
\end{figure}

We note that the other quantities depend on
$E_c$ under the regularization~(\ref{regularization_g}),
while the cutoff dependence of
the order parameter can be removed.
Figure~\ref{mu_Ecdep} shows that
the chemical potential and the eigenvalues for the
eigenfunctions with 0 to 3 nodes at $l=0$
depend on $E_c$.
In general, the energy spectrum varies with $E_c$.
As shown in Fig. 8, increasing $E_c$ appears to
eliminate the cutoff dependence.
However, if the energy cutoff $E_c$ is set too high,
the Higgs mode does not appear in the numerical solution of
the integral equations (\ref{mode_equation_radial}).
This indicates that although the renormalization-based correction
of the interaction strength (\ref{regularization_g})
is effective to derive the order parameter,
it is inadequate to calculate the Higgs mode frequency.
This is discussed in Appendix~\ref{cutoff_dependence_Higgs}.

\section{Cutoff dependence of Higgs mode}
\label{cutoff_dependence_Higgs}

\begin{figure}
    \includegraphics[width=\linewidth]
    {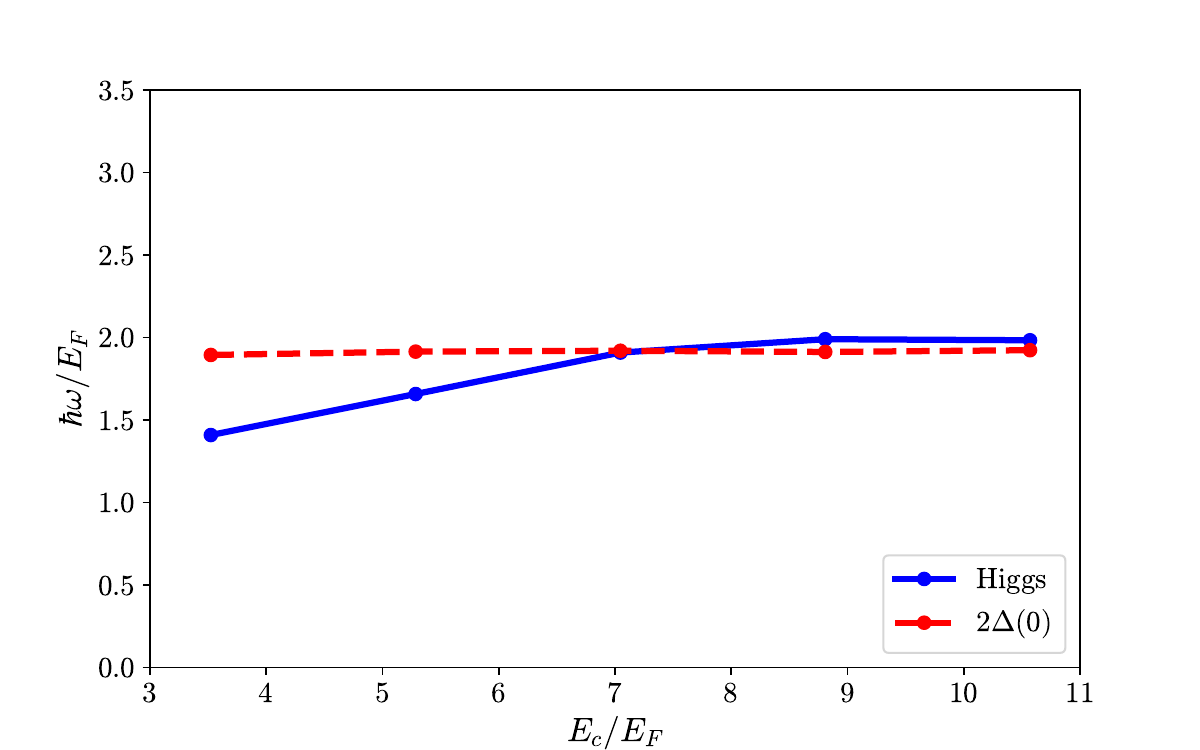}% Here is how to import EPS art
    \caption{
        The energy cutoff dependence of the Higgs mode.
        The values of the Higgs mode are
        plotted for $1/(a_Sk_F)=-0.6,
        1/(a_{\text{ho}}k_F)=0.15$ and $T/T_F=0$
        as a function of the cutoff $E_c$.
        The blue and red line are
        the frequency of the Higgs mode and
        twice the central value of the order parameter,
        respectively.
    }
\label{Higgs_Ecdep}
\end{figure}

We see the cutoff dependence of the Higgs mode
with the rectification (\ref{regularization_g}).
Figure \ref{Higgs_Ecdep} shows that the frequency
of the Higgs mode converges approximately twice the
central value of the order parameter
for energy cutoffs $E_c/E_F \geq 7$
for $1/(a_Sk_F)=-0.6, 1/(a_{\text{ho}}k_F)=0.15$
and $T/T_F=0$.
On the other hand, explicit results of the Higgs mode
frequency cannot be obtained with $E_c/E_F \geq 11$,
while the BdG equation~(\ref{radial_BdG}) provides
the converged order parameter.
This implies that excessive regularization prevents
the stable numerical evaluation of
Eq.~(\ref{mode_equation_radial}).

\section{Verification of NG mode}
\begin{figure}
    \includegraphics[width=\linewidth]
    {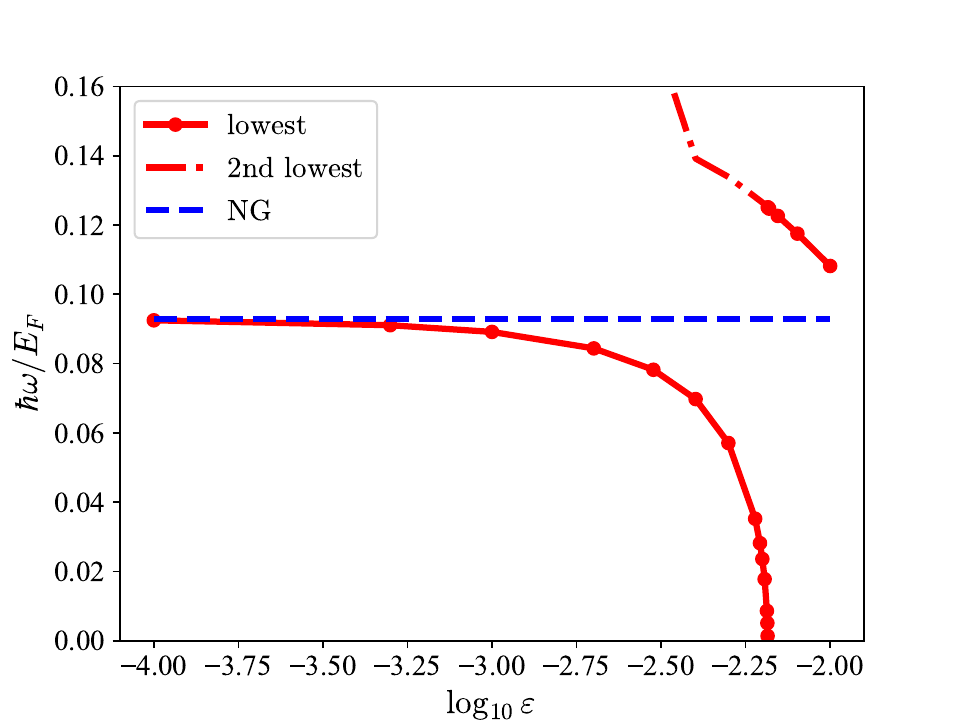}
    % Here is how to import EPS art
    \caption{
        The error dependence of the NG mode.
        The frequencies of the NG mode are plotted for
        $1/(a_Sk_F)=-0.6, 1/(a_{\text{ho}}k_F)=0.15,
        T/T_F=0$ and $E_c/E_F=5.29$ as a function of
        the breaking parameter $\varepsilon$.
        The blue line is the value of the NG mode
        obtained by Eq. (\ref{characteristic_eq_NG}).
        The solid and dash-dotted red line are
        the lowest and second lowest solutions
        of Eq.~(\ref{characteristic_eq_NG_break}),
        respectively.
    }
\label{NG_break}
\end{figure}

We verify the emergence of the NG mode
in an isotropic harmonic trapped system
by calculating the
following characteristic equation
\begin{equation}
    \operatorname{det}
    \left[
        \mathbf{I}
        -
        \mathbf{M}_{22}(\omega)
    \right]
    =   0.
\label{characteristic_eq_NG}
\end{equation}
However, it does not incorporate
an exact zero mode.
An oscillation frequency of the phase
fluctuation is not exactly equal to zero.
To determine whether this result
is due to numerical errors,
we introduce a breaking term into
numerical calculations
\begin{equation}
    \operatorname{det}
    \left[
        (1 - \varepsilon) \mathbf{I}
        -
        \mathbf{M}_{22}(\omega)
    \right]
    =   0,
\label{characteristic_eq_NG_break}
\end{equation}
where $\varepsilon$ is the breaking parameter.
Figure \ref{NG_break} shows that
the lowest solution $\omega$ becomes zero
where $\varepsilon$ is about
$6.53 \times 10^{-3}$.
When $\varepsilon$ is set to greater than
$6.53 \times 10^{-3}$, the smallest solution
disappears, and the second smallest
becomes the smallest one.
Thus, the NG mode emerges even when Eq. (\ref{mode_equation})
is applied to inhomogeneous systems.

%\nocite{*}

\bibliography{main}% Produces the bibliography via BibTeX.

\end{document}